# The Status of CMS

Joseph R. Incandela

*Physics Department, University of California, Santa Barbara, California, 93106, USA*

**Abstract.** After a brief overview of the Compact Muon Solenoid (CMS) experiment, the status of construction, installation and commissioning is described. Very good progress has been achieved in the past year. Though many significant challenges still lie ahead, CMS should be ready for recording data from first collisions in the Large Hadron Collider (LHC) accelerator complex at CERN in late 2007.

## INTRODUCTION

Although the Standard Model (SM) of particle physics has proven to be extremely successful in its description of nature at the highest energies and smallest distance scales accessible to physics to date, it is by no means a complete theory and leaves many important questions unanswered. A key tenet of the SM is the necessity of electroweak symmetry breaking (EWSB) for which the Higgs mechanism is presumed, but not yet proven, to be responsible. The nature of EWSB is expected to be manifested at the TeV energy scale that will be accessed by the Large Hadron Collider (LHC) complex. There are also many alternative possible sources of EWSB such as Technicolor, and there exists the possibility that the true source of EWSB may differ from anything that has yet been considered. Beyond EWSB, the LHC program may yield discoveries that could elucidate the nature of a unified theory underlying the SM. Examples include the discovery of a spectrum of supersymmetric partners of the usual cast of SM particles, or evidence of extra dimensions resulting from gravity at the TeV scale. The Compact Muon Solenoid (CMS) at the LHC will have very broad capabilities allowing exploration of these phenomena in proton-proton interactions at a center-of-mass energy of 14 TeV – seven times the highest energy achieved previously at the Fermilab Tevatron. In addition, heavy-ion beams at energies over 30 times higher than previously accessible will be provided by the LHC and their collisions will be studied by CMS.

## OVERVIEW OF THE CMS DETECTOR [1]

A schematic drawing of CMS is shown in Figure 1. The total weight of the apparatus is 12500 tons. The detector, which is cylindrical in shape, has length and diameter of 21.6 m and 14.6 m, respectively. The overall size is set by the muon tracking system which in turn makes use of the return flux of a 13 m-long, 5.9 m-diameter, 4 Tesla superconducting solenoid. This rather high field was chosen to facilitate the construction of a compact tracking system on its interior while also allowing good muon tracking on the exterior without the need for excessive demands being placed on muon-chamber resolution and alignment. The return field saturates 1.5 m of iron into which is interleaved four muon tracking stations. In the central region (pseudorapidity range $|\eta| < 1.2$) the neutron induced background, the muon rate and the residual magnetic fields are all relatively small, while in the forward regions ($1.2 < |\eta| < 2.4$) all three quantities are relatively high. As a result, drift tube (DT) chambers and cathode strip chambers (CSC), are used for muon tracking in the central and forward regions, respectively. Resistive plate chambers (RPC) with fast response and good time resolution but coarser position resolution are used in both regions for timing and redundancy.

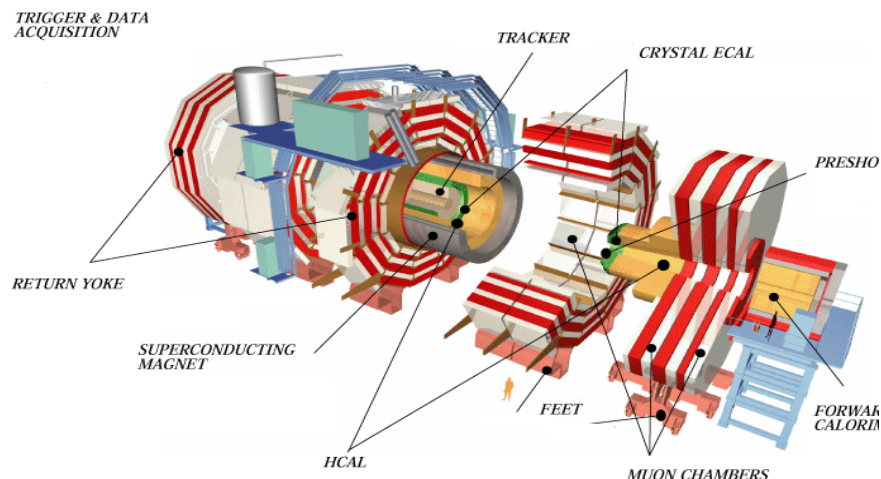

Figure 1. Schematic drawing of the CMS detector with a portion removed to reveal the various subsystems as indicated. The detector is divided into 5 barrel Yoke sections, and 3 endcap yokes per end. The HF is mounted on the stand at far right.

The solenoid is large enough to house the inner tracker and the calorimetry. The sensing elements of the EM calorimeter (ECAL) are PbWO$_4$ (lead tungstenate) crystals. The crystals can tolerate a radiation dose of 10 Mrad and have short radiation ($X_0$ = 0.89 cm) and Molière (2.2 cm) lengths which are ideal for the design of a compact calorimeter with fine granularity. They are also fast, emitting 80% of all scintillation light within the 25 ns spacing between proton bunch crossings at the LHC. This light is detected by Si avalanche photodiodes (APD) and vacuum phototriodes (VPT) in the barrel and endcap regions, respectively. Due to their sensitivity to temperature change, both the crystals and the APD require a temperature stability of ~0.1 C to take full advantage of the excellent inherent energy resolution of the crystals. A preshower system is installed in front of the endcap ECAL to help detect $\pi^o \rightarrow \gamma\gamma$.

The CMS central hadronic calorimeter (HCAL) is outside of the ECAL but also inside the magnet coil. As a result of the latter, the absorber material in the HCAL is brass. A fraction of high energy hadronic showers extend beyond the HCAL and are taken into account by means of the signals retrieved from an additional layer of scintillators comprising the hadron outer (HO) detector that line the outside of the coil. The active elements consist of plastic scintillator tiles with embedded wavelength-shifting fibers that are spliced to long attenuation length fibers. The latter carry the light to hybrid photo-diodes which provide signal amplification in high axial magnetic fields. These technologies make it possible to build the HCAL with almost no un-instrumented cracks or dead areas in $|\eta|$. The endcap hadron calorimeter (HE) uses the same technology as the HCAL and covers the pseudorapidity region $1.3 < |\eta| < 3$ while the region $3 < |\eta| < 5$ is covered by the Fe and quartz-fiber Hadron Forward (HF) calorimeter. Cerenkov light emitted in the quartz fibers is detected by fast photomultipliers. The HF technology is ideal for the dense jet environment typical of this region as it leads to narrower and shorter hadronic showers. Calorimeter coverage to $|\eta| < 5$ is useful for reducing the uncertainty on missing transverse energy.

The CMS tracker occupies a cylindrical volume of length 5.8 m and diameter 2.6 m. The outer portion of the tracker is comprised of 10 layers of silicon microstrip detectors and the inner portion is made up of 3 layers of silicon pixels. Silicon provides fine granularity and precision in all regions for efficient and pure track reconstruction even in the very dense track environment of high energy jets. The three layers of silicon pixel detectors at radii of 4, 7 and 11 cm provide 3D space points that are used to seed the formation of tracks by the pattern recognition. The 3D points also enable measurement of the impact parameters of charged-particle tracks with a precision of order 20 µm in both the r-$\phi$ and r-z views. The latter allows for precise reconstruction of displaced vertices to yield efficient b tagging and good separation between heavy and light quark jets.

In regard to performance, the CMS experiment is designed for:
- Good µ and charged particle tracking with momentum resolution over a wide range of momenta in $|\eta| < 2.5$.
- Relatively high efficiency heavy flavor and $\tau$ jet tagging with low rates for tagging light quark jets.
- Very good e and $\gamma$ energy resolution in the region $|\eta| < 2.5$ and good separation of $\gamma$'s and e's from $\pi$'s.

- The ability to determine the direction of photons and/or identify the relevant primary interaction vertex.
- Good missing $E_T$ and dijet mass resolution with fine lateral segmentation ($\Delta\eta \times \Delta\phi < 0.1 \times 0.1$) in HCAL.

## STATUS OF CMS SUBSYSTEMS

The realization of the CMS apparatus is proceeding well. The past year has been a particularly important one in terms of the achievement of significant milestones such as the completion of all tracker modules and subassemblies and the delivery of the majority of ECAL crystals. Nevertheless, there remain significant challenges in the installation and commissioning of CMS in time for first collisions in late 2007. In this section, the status of each major subsystem, as of August 2006, is presented starting with the interaction point and surrounding infrastructure ("Point 5" in Cessy, France), and then working inward through the various detector subsystems.

Civil Engineering work at Point 5 in Cessy, France has finished. The counting room is ready for sub-detector readout crate installation and the experimental cavern (UX5) is ready to receive detector elements. For ease of assembly, installation and maintenance, the Barrel yoke is sectioned into 5 ring-sections and each Endcap yoke into 3 disk-sections. The Hadronic Forward (HF) calorimeters are to be pushed outside of the yoke to allow movement of the endcaps along the beam-pipe. This modularity, as represented in Figure 1, allows for the assembly of major portions of the detector in the surface building (SX5) where there are fewer space constraints. They will then be lowered into their final positions in the cavern. All the CMS sub-detectors will be essentially commissioned as large systems, including electronics, power and control systems, on the surface before they are lowered into the experimental cavern. Lowering CMS into the experimental cavern requires 15 large-lift operations.

The CMS assembly started several years ago in SX5. The barrel and endcap yoke were assembled first; the HCAL next, followed by installation of muon chambers on the endcap yoke and inside the barrel wheels. The solenoid coil has been assembled installed and very recently operated at full field.

## The Muon System

Each Muon Endcap (ME) system contains 234 CSC. Currently, 432 of the 468 endcap CSC have been installed on the magnet yoke disks and are now being commissioned with cosmic rays. Chambers as they appear after installation on two disks are shown in Figure 2. Each trapezoidal endcap CSC chamber has 6 gas gaps containing a plane of radial cathode strips and a plane of anode wires parallel to the longest edge of the trapezoid and so, roughly perpendicular to the strips. The spatial resolution provided by each chamber ranges from 100 μm in Station 1 to roughly 150 μm in Stations 2 thru 4. Wire signals are fast and are used in the Level-1 Trigger though they have coarser position resolution.

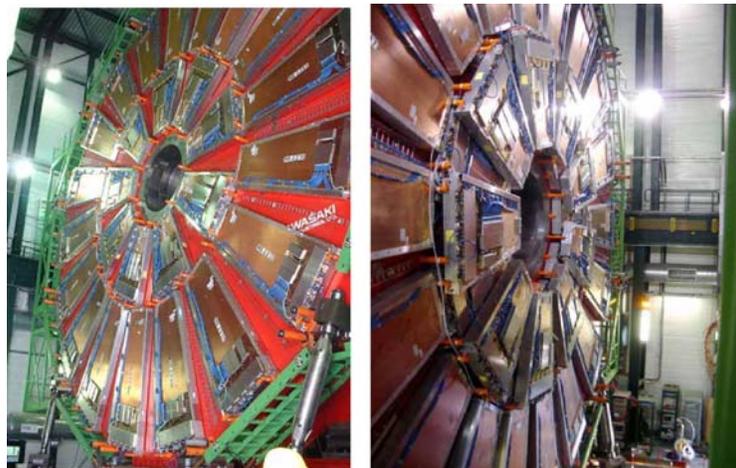

**FIGURE 2.** Cathode Strip Chambers (CSC) installed on endcap yokes.

The manufacture of Barrel DT chambers is complete and 146 of 266 DT have been installed. The chambers installed in barrel yokes (YB) are organized in 4 stations as seen in Figure 3. Each DT chamber is piggy-backed by one or two RPC. The chambers are staggered from station to station so that a high-$p_T$ muon near a sector boundary crosses at least 3 stations. The chambers consist of twelve planes of aluminum drift tubes; four r-φ measuring planes are placed above, and four below, a group of four z-measuring planes. Each station gives a muon vector in space with a precision of <100 μm in position and <1 mrad in direction. At present, 278 of 460 RPC have been installed. The forward RPC system covers |η| < 2.1 but those chambers at |η| > 1.6 have been staged for later installation.

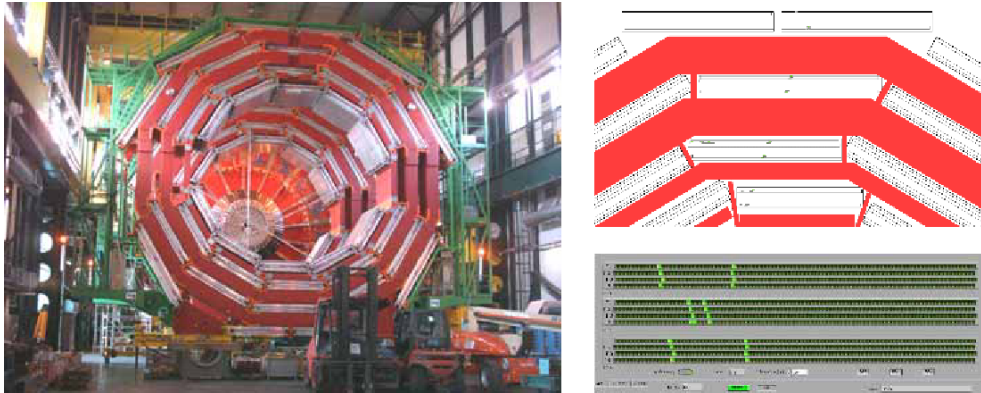

**FIGURE 3.** Left: DT stations installed on a barrel yoke; Right: display of a cosmic μ passing through 4 stations (top) and the 12 layers of one station (bottom) containing 4 z-planes sandwiched between 2 groups of 4 r-φ planes.

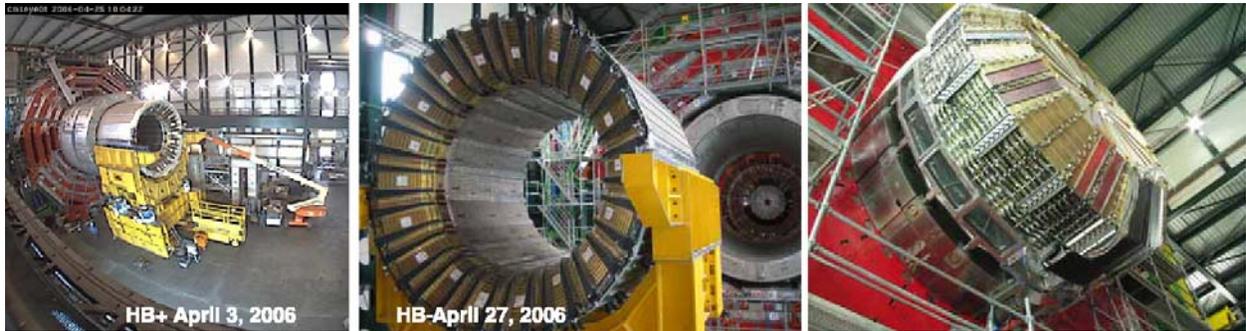

**FIGURE 4.** The CMS Hadron Calorimeter (HCAL): the two barrel sections (denoted HB+ and HB-) are shown during their installation in April, 2006. At right is one of the end caps (HF).

## The Hadronic Calorimeter

All HCAL module types [HB (barrel), HE (endcap), HO (outer) and HF (forward)], including absorber and optics, are completed. Photodetectors and electronics have been installed and a comprehensive calibration of HCAL using Co-60 sources has been completed. HF will be the first sub-detector to be lowered into CX5. This is expected to occur in October, 2006. The HB and HE are fully installed as shown in Figure 4. The HF has been calibrated to ~ 5%, HE and HB to ~ 4%. The HE and HB have also been run globally with muons. HCAL Trigger and data acquisition (DAQ) have been tested in cosmics. HCAL slow controls are fully operational and data quality monitoring (DQM) is now under development and partially up and running.

## The Electromagnetic Calorimeter

The barrel section (EB) has an inner radius of 129 cm. It is made up of 36 identical supermodules, each covering half the barrel length and corresponding to a pseudorapidity interval of 0 < |η| < 1.479. The crystals have a front face cross-section of ~ 22 x 22 mm$^2$ and a length of 230 mm, (corresponding to 25.8 $X_0$) as seen in Fig. 5. When installed, they are arrayed in an η–φ grid with the axis of each crystal projecting near to, but not directly at, the nominal center point of the CMS detector. The endcaps (EE) cover a pseudorapidity range of 1.479 < |η| < 3.0. They have a front face cross section of 28.6 x 28.6 mm$^2$ and a length of 220 mm (24.7 $X_0$). Like the barrel crystals, those in the endcap project to points near to the detector center point, but they are arranged in an x-y grid. At the time of this writing, some 54600 out of 61200 barrel crystals have been delivered. Barrel and endcap crystal deliveries are expected to be complete in Feb. 2007 and Jan. 2008, respectively.

Construction of the 36 supermodules (SM), each of which contains 1700 crystals, is well underway. Each of the SM is tested for one week after assembly followed by another week of operation in cosmic rays to obtain a relative calibration of all crystals. At the time of this writing, 30 bare SM have been constructed, of which 22 have been fully integrated with electronics and crystals and 18 have been calibrated with cosmics and 4 in test beam. The barrel ECAL will be ready for the pilot run in 2007.

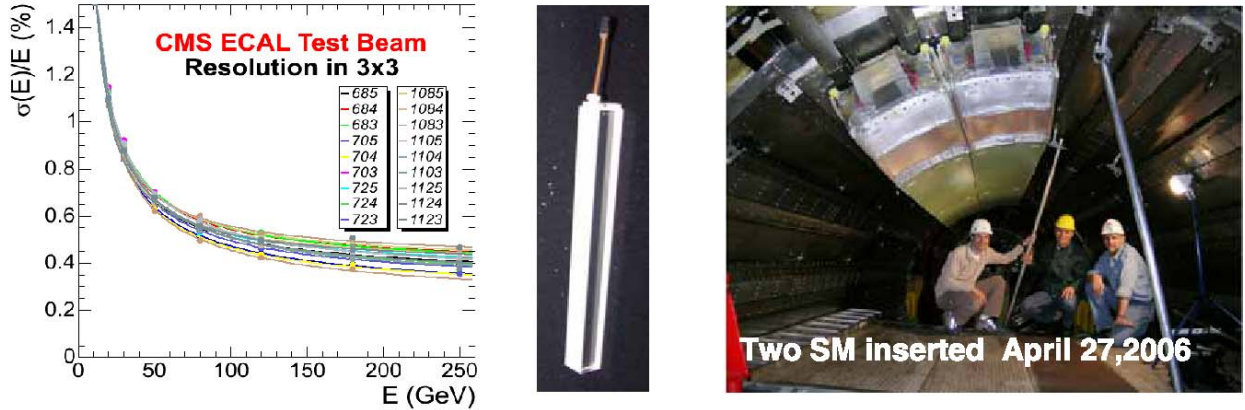

**FIGURE 5.** The CMS Electromagnetic Calorimeter (ECAL) (from left to right): the performance of 18 crystals in test beam; photo of an individual PbWO$_4$ crystal and the first two supermodules just after installation in April, 2006.

The readout of the signals from the crystals in the ECAL is designed to obtain optimal energy resolution over a large dynamic range. The signal is amplified by a multi-gain preamplifier and shaped to peak at ~ 50 ns. The shaped signal is sampled and digitized at 40 MHz in one of three 12-bit ADCs corresponding to different energy ranges. A dynamic range of over 15 bits is thus achieved. The noise performance for completed SM is close to the original specification of 40 MeV/channel. The energy resolution parameterized as a function of energy in GeV with test beam data for a completed SM is:

$$\left(\frac{\sigma}{E}\right)^2 = \left(\frac{3\%}{\sqrt{E}}\right)^2 + \left(\frac{0.16\%}{E}\right)^2 + (0.3\%)^2$$

## The Inner Tracker

As noted earlier, there are 3 layers of hybrid pixel detectors at radii of 4, 7, and 11 cm in the barrel region. The size of the pixels is 100 x 150 μm$^2$. Outside these layers are 10 layers of Si microstrip detectors distributed roughly uniformly in radius between 20 and 115 cm. The forward region has two pixel and nine micro-strip layers in each of the two endcaps. These are arrayed in circular disks. The pixel system has a sensitive area of 1 m$^2$, while the silicon strips have area of 207 m$^2$, providing coverage for $|\eta| < 2.4$. All told, there are 66 M pixels and 9.6 M silicon strips.

Construction of the Si strip tracker occurs in 3 phases: module production, assembly of modules into supermodules (known as Rods in the Tracker Outer Barrel (TOB), Shells in the Tracker Inner Barrel (TIB), and Petals in the Tracker End Caps (TEC)), installation of supermodules into the TOB, TIB and TEC and their integration in the Tracker Support Tube. The first phase was completed in spring, the second is complete for the TOB and TIB and nearly so for the TEC, and the third is well underway with completion expected by the end of 2006. Figure 6 includes photographs of one half of the TIB, one of the two TEC systems, and one end of the TOB during assembly at the Tracker Integration Facility (TIF) in building 186 at CERN in late July 2006. Installation of all Petals for one the TEC systems was completed on Sep. 5, 2006. Figure 7 shows the typical noise performance for an entire TOB cooling segment during integration. The uniformity is outstanding and there are more than 99.9% of channels functioning. Similar results are found for the TEC and TIB.

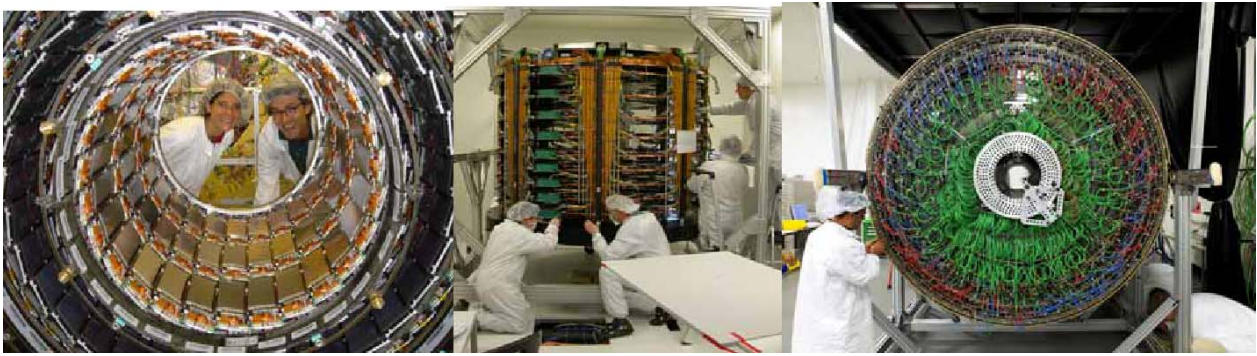

**FIGURE 6.** CMS silicon tracker subsystems in the Tracker Integration Facility at CERN are shown with some of the integration and testing personnel; from left to right the TIB, TEC and TOB.

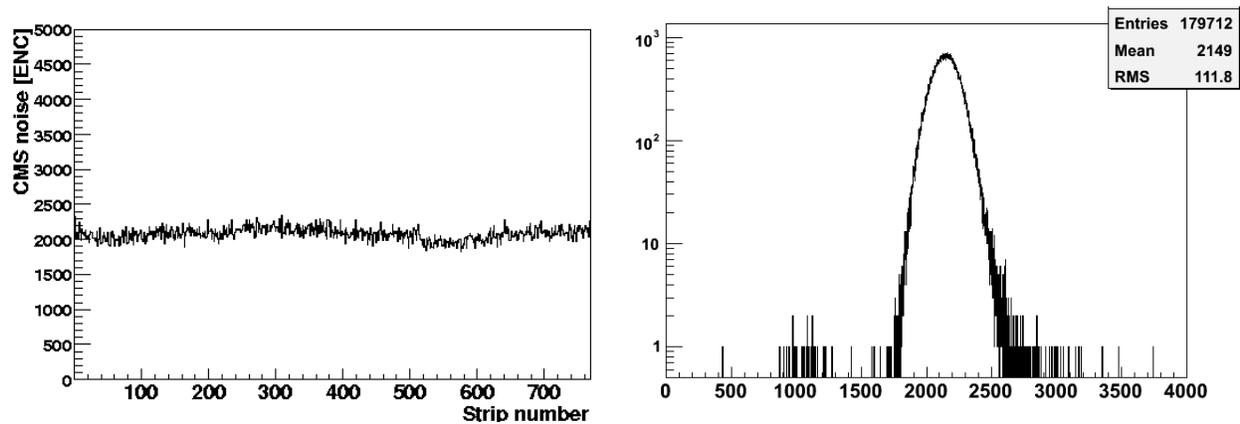

**FIGURE 7.** Noise performance of the CMS TOB: Left; noise on all channels of a typical TOB module, Right; the distribution of the average noise for all 179k channels of a TOB cooling segment. A handful of channels clustered at ~1000 ENC are those for which the second sensor is disconnected because of a faulty strip. A minimum ionizing particle produces a minimum charge deposition of order ~45k electrons at normal incidence, (Si thickness is 500 μm).

The pixel detector components are now in final production. The system is built up of various subassemblies: *Plaquettes*, as seen in Fig. 8, are sensors bump-bonded to readout chips and attached to very high density Kapton interconnect cables. Plaquettes are then assembled into *panels*. At present, somewhat less than half of all plaquettes have been assembled. Integration of plaquettes into panels has just started. A small number of panels will be installed for the 2007 pilot run. The entire pixel detector is expected to be complete and installed for the first physics run in 2008.

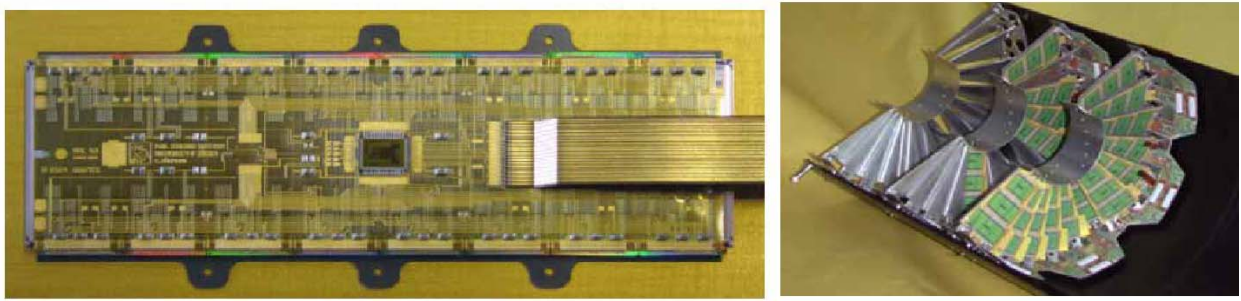

**FIGURE 8.** Left: A CMS pixel detector plaquette. Right: mock-up of a portion of the end-cap pixel detectors.

## The Level-1 Trigger and Data Acquisition Systems

The trigger system is well into production with many components already completed. Components are exercised with other trigger and detector electronics systems in the Electronics Integration Center at building 904 (CERN-Prevessin). Integration tests are run in which detector primitives are generated and used to feed the trigger and DAQ.

Final system components were also used in data taking during the magnet test at SX5 this summer in which the magnet was operated at its operating strength of 4 T. The magnetic field was mapped and a combined test of all sub-detectors was performed using cosmic rays. The data from this test are currently being analyzed.

## PREPARATION FOR PHYSICS

The software for the experiment has been used to perform detailed simulations of the detector response, to develop sophisticated reconstruction algorithms and to perform detailed studies of a broad range of physics topics. This work is detailed in Technical Design Reports (TDR), several of which are already public including the Computing TDR [2] and the Physics TDR Volume 1 [3]. Volume 2 of the Physics TDR [4], containing detailed physics performance studies, was completed this summer. Many of the important results are reported elsewhere in this conference.

Recently CMS determined that it would be advantageous to restructure the software framework in preparation for CMS data-taking in order to implement calibration and alignment strategies, standardize the reconstruction modules, and facilitate interactive analyses. The new framework, known as CMSSW is now being widely used, both in detector operation and offline studies of cosmic and test beam data and simulated data.

Later this year a Computing, Software and Analysis Challenge (CSA2006) is planned. This will test the complete system, from (simulated) raw data to final analysis at Tier-1 and Tier-2 centers.

## CONCLUSIONS

After many years of R&D, prototyping and construction, the majority of the CMS detector components are now finished. As was true for other LHC experiments, this has entailed an industrial scale of production of very high precision components that is unprecedented in the history of high energy physics. Installation and commissioning is underway and proceeding well. CMS is expected to be ready for beam in the latter half of 2007. The CMS detector should be capable of discovering the true nature of the TeV energy scale which could provide clues to answering some of the biggest questions in physics.

## ACKNOWLEDGEMENTS

I thank the organizers of the Hadron Collider Physics 2006 conference for their hospitality and for providing an excellent venue and environment for this important event. The CMS experiment is the result of the creativity and intense efforts of thousands of scientists, engineers and technicians worldwide, who are gratefully acknowledged.

I would also like to thank D. Acosta, E. Auffray, J. Butler, M. Della Negra, J. Freeman, D. Green, R. Rusack, W. Smith, S. Tkaczyk, and T. Virdee for providing up-to-date information, R. Tenchini for useful comments and H. Incandela for help in preparing this document.